# Digital Forensic Readiness Implementation in SDN: Issues and Challenges

[1]**Nickson M. Karie[*] and [2]Craig Valli[**]**
[1,2]Cyber Security Cooperative Research Centre, Australia
[1,2]Security Research Institute, Edith Cowan University, WA, Australia.
nickson.karie@cybersecuritycrc.org.au[*],
c.valli@ecu.edu.au[**]

**Abstract:** The continued evolution in computer network technologies has seen the introduction of new paradigms like Software Defined Networking (SDN) which has altered many traditional networking principles in today's business environments. SDN has brought about unprecedented change to the way organisations plan, develop, and enact their networking technology and infrastructure strategies. However, SDN does not only offer new opportunities and abilities for organisations to redesign their entire network infrastructure but also presents a different set of issues and challenges that need to be resolved. One such challenge is the implementation of Digital Forensic Readiness (DFR) in SDN environments. This paper, therefore, examines existing literature and highlights the different issues and challenges impacting the implementation of DFR in SDN. However, the paper also goes further to offer insights on the different countermeasures that organisations can embrace to enhance their ability to respond to cybersecurity incidents as well as help them in implementing DFR in SDN environments.

## 1. Introduction

Software-defined networking (SDN) can be defined as "an approach to networking that uses software-based controllers or application programming interfaces to communicate with the underlying hardware infrastructure and direct traffic on a network" (VMware, 2021). According to Kandoi and Antikainen, (2015) SDN has recently gained significant momentum in many organisations. Amin, Reisslein, and Shah, (2018) adds that SDN gives organisations the power to decouple the control plane from the data plane of forwarding devices thus simplifying network management and control, making computer networks agile and flexible. However, Mugitama, Cahyani, and Sukamo (2020) state that since its inception, the SDN architecture which mainly abstracts different, distinguishable layers of a network was not designed with a focus on network security. For this reason, the centralized control nature of SDN introduces some security vulnerabilities that can be exploited, for example, using Denial of Service (DoS) attacks to cause packet overload or race condition on the controller.

Another known security vulnerability at the SDN controller level is the topology poisoning attack. This attack utilizes spoofed packets and exploits the Link Layer Discovery Protocol (LLDP) packets in the network. Topology poisoning attack is one among many other malicious activities and security vulnerabilities that makes Digital Forensic Investigation (DFI) a challenging process in SDN environments. Also, the lack of standardised approaches specifically designed to help forensic investigators in SDN environments adds to the complexity of conducting digital forensic investigations in SDN.

As organisations continue to embrace SDN, many of them are likely to be targeted or abused by malicious actors to facilitate malicious cyber activities (Kebande, et.al, 2020). The ability to execute malicious activities in SDN environments to cause harm or abuse makes SDN forensics an increasingly important process and reinforces the importance of implementing DFR in organisations (Karie and Karume, 2017).

This paper, therefore, examines existing literature and highlights the different issues and challenges impacting the implementation of DFR in SDN environments. Further, this paper highlights different countermeasures that organisations can embrace to enhance their ability to respond to cybersecurity incidents as well as help them in implementing DFR in SDN environments. Note at this point also that this paper utilised purposive sampling research methodology where the authors relied on their own judgment to choose the literature examined and used in this study.

As for the remaining part of this paper: Section 2 introduces the literature review as background while Section 3 presents related work. Thereafter, Section 4 discusses the issues and challenges impacting the implementation of DFR in SDN. Section 5 presents different countermeasures that organisations can embrace to enhance their ability to respond to cybersecurity incidents as well as help them in implementing DFR in SDN environments before concluding the paper in Section 6 and makes mention of the future work.

## 2. Background

This section provides a literature background on the following areas: digital forensics, Digital Forensic Readiness (DFR), and Software-Defined Networking (SDN). Digital forensics helps to understand the scientific process used for conducting digital forensic investigations while DFR is discussed as a way that can help organisations record activities and data in such a manner that the records are sufficient in their extent for subsequent forensic purposes thus minimizing digital forensic investigation costs (Mohay, 2005). Finally, SDN is discussed to help understand the concept of physical separation of the network control plane from the forwarding plane and how the control plane is used to control different devices in the network.

### 1.1 Digital Forensics

Continued malicious activities in the cyberspace makes digital forensics an essential process for investigating and prosecuting cybercriminals misusing digital devices such as computer systems, network devices, mobile devices, and storage devices (Selamat, Yusof, and Sahib, 2008). For this reason, digital forensics can be described as a scientific process of investigation that deals with extracting and analysing digital artefacts from digital devices. According to Ademu, Imafidon, and Preston (2011), digital forensics provides tools, techniques, and scientifically proven methods that can be used to acquire and analyse digital artefacts. The acquired or extracted digital artefacts can be used to reconstruct events for purposes of creating a hypothesis that can be useful in a court of law or any civil proceedings (Karie and Kebande, 2016).

Any successful digital forensic investigation process involves a thorough forensic examination of digital artefacts by forensic analysts with the primary objective being to unearth information that will assist practitioners and law enforcement agencies in the presentation of digital evidence in a court of law or any civil proceedings. In this regard, Karie and Kebande (2016) add that digital forensics thus expands simply from the crime scene through digital forensic analysts to the courtroom. This also implies that, if the outcome of an investigation is to be presented in a court of law as evidence, the digital forensic investigation process must always adhere to some important scientifically proven and accepted methods or processes that must be considered and taken (Köhn, Olivier, and Eloff, 2006). For a comprehensive reading on digital forensic investigation processes, the reader is advised to consult Valjarevic, and Venter, (2015), Valjarevic and Venter, (2012), and Köhn, Olivier, and Eloff, (2006). The next subsection presents the concept of digital forensic readiness.

### 1.2 Digital Forensic Readiness (DFR)

Based on the definitions and description of digital forensics, the digital forensic investigation process is usually employed as a post-event response to cybersecurity incidents (Rowlingson, 2004). However, for an organisation to benefit from the ability to gather and preserve digital evidence before an incident occurs, DFR is inevitable. DFR can be defined as "the ability of an organisation to maximize its potential to use digital evidence whilst minimizing the costs of an investigation" (Rowlingson, 2004). From this definition, DFR allows an organisation to regain control and limit the damage and costs from any security incident (KPMG, 2015) as well as demonstrate due diligence with regulations, conduct digital forensic investigations, and produce digital artefacts that can be used in a court of law or civil proceedings as evidence (DFMAG, 2020). This also implies that DFR can help organisations record activities and data in such a manner that the records are sufficient in their extent for subsequent forensic purposes thus minimizing digital forensic investigation costs (Mohay, 2005).

Antonio and Labuschagne (2013) add that a carefully considered and planned legally contextualized DFR strategy can provide organisations with an increased ability to respond to security incidents while maintaining the integrity of the evidence gathered and keeping investigative costs low. DFR can also help organisations with quicker recovery, improved business continuity, and compliance, as well as an improved success rate in legal actions by having available the collected digital artefacts for use as evidence (Andre, 2014). From this DFR description, the authors agree with the sentiments echoed by Elyas, Ahmad, Maynard, and Lonie, (2015) that implementing DFR in SDN environments can help organisations to comply with their legal, contractual, regulatory, security, and operational obligations. The concept of SDN is covered in the next subsection.

### 1.3 Software-Defined Networking (SDN)

SDN is an emerging networking technology offering centralised control of computer networks (Mugitama, Cahyani, and Sukamo, 2020) that allows network operators to dynamically configure and manage their infrastructures. The primary difference between SDN and traditional networking is that SDN is software-based, while traditional networking is hardware-based (VMware, 2021). Because of the programmability of SDN, organisations can experience more advantages of having SDN than traditional networking techniques. Some of the primary advantages of SDN are increased control with greater speed and flexibility, customizable network infrastructure as well as robust security.

The goal of SDN however as stated by Rosencrance, English, and Burke, (2020) is to "improve network control by enabling enterprises and service providers to respond quickly to changing business requirements". This is backed up by the fact that in SDN, "a network engineer or administrator can shape traffic from a centralized control console without having to touch individual switches in the network. Besides, a centralized SDN controller will direct the switches or routers to deliver network services wherever they are needed, regardless of the specific connections between a server and devices" (Rosencrance, English, and Burke, 2020). Figure 1 shows a simplified SDN architecture.

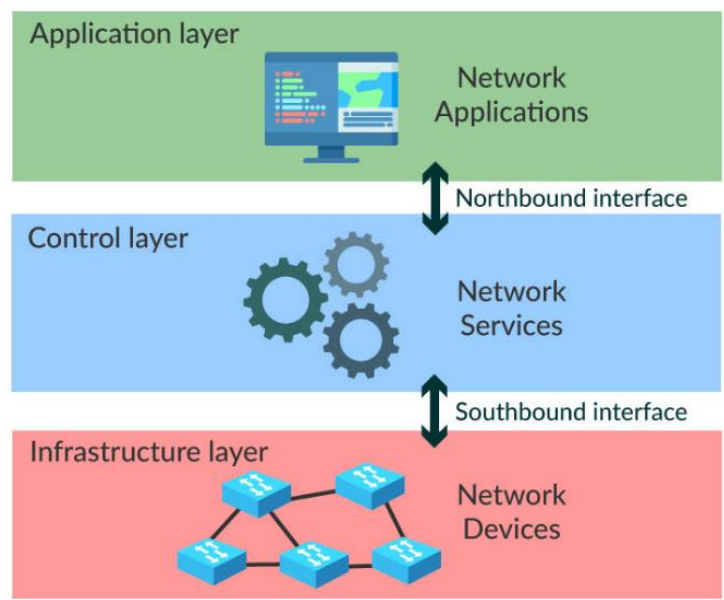

Figure 1: Software-Defined Network (SDN) Architecture
Source: (https://electronicsguide4u.com/sdn-network-software-defined-network-openflow-protocol-what-is-sdn/)

Infer from Figure 1 that SDN can be divided into three-tier architecture with three primary layers. Communication through interfaces between the different tiers or layers is usually handled by the northbound and southbound interfaces as shown in Figure 1. The three basic layers of SDN are brief explained in the following subsections.

*1.3.1 Application Layer*

The application layer as shown in Figure 1 hosts the SDN applications or network applications which are programs designed to perform different tasks. SDN applications explicitly, directly, and programmatically communicate their network requirements and desired network behaviour to the SDN controller via a northbound interface (ONF, 2013). The northbound interface (NBI) in this case serves as a communication link between SDN applications and the SDN controller. Again, the NBI provides an abstract network view that enables the direct expression of network behaviour and requirements. Examples of SDN applications include networking management, analytics as well as other business applications used mostly in data centres (SDx, 2015).

*1.3.2 Control Layer*

The control layer oversees the network intelligence and hosts the control logic for managing the network services. The SDN controller which also acts as the brain of the entire network manages and manipulates flow entries to and from multiple devices (Aric, 2020). As seen from Figure 1 the control layer houses the SDN controller which is a logically centralized entity. According to ONF (2013), some of the functions of the control layer include but are not limited to translating the requirements from the application layer to the data paths and providing SDN applications with an abstract view of the network (ONF, 2013). Note that the SDN data path is a software-based representation of a network device in SDN environments (Kemp,2020) that exposes visibility and uncontested control over its advertised forwarding and data processing capabilities (ONF, 2013).

*1.3.3 Infrastructure Layer*

The infrastructure layer which also forms the physical layer of the network consists of various network devices (network switches and routers) or networking equipment (Aric, 2020) and forms the underlying network to forward network traffic. The network devices found in the infrastructure layer are responsible for handling network packets based on the rules provided by the SDN controller. The next section presents related works.

## 3. Related Works

Several related research work from different researchers exists and have made valuable contributions to the study presented in this paper. In this section, a summary of some of the most prominent efforts is presented.

To begin with, Munkhondya, Ikuesan, and Venter, (2019) proposed a DFR approach for potential evidence preservation in SDN. However, their research focused on developing a proactive mechanism for the identification, handling, collection, and preservation of digital artefacts in SDN. Also, their proposed mechanism was meant to integrate the DFR approach to the acquisition and preservation of volatile artefacts. Their research however did not capture the specific issues and challenges impacting the implementation of DFR in SDN environments as is the case of this current study.

Another effort by Lagrasse, Singh, Munkhondya, Ikuesan, and Venter, (2020) proposed "a proactive DFR framework for SDN with a trigger-based automated collection mechanism. Their proposed mechanism integrated an intrusion detection system and an SDN controller". Their research mainly concentrated on how best to collect digital artefacts in SDN environments without regard to the issues and challenges impacting the implementation of DFR in SDN. The current study in this paper, however, investigates this research gap by discussing the issues and challenges impacting the implementation of DFR in SDN environments.

Sezer, et.al. (2013) raised the question of how to achieve a successful carrier-grade network with SDN. In their research, however, Sezer, et.al. (2013) only focus on the challenges of network performance, scalability, security, and interoperability with the proposal of potential solution directions. Their research had no reference to the issues and challenges impacting the implementation of DFR in SDN environments as is the case discussed in this current paper.

Efforts by Munkhondya, Ikuesan, and Venter, (2020) presented a case for a dynamic approach to DFR in SDN environments. Their research argues that "the DFR approach mostly employed has been limited to static potential digital evidence collection which contrasts the dynamic nature of SDN environments" (Munkhondya, Ikuesan, and Venter, 2020). They then go ahead and discuss the pitfalls of the static DFR approach which underscores the need for a dynamic DFR approach. The current paper however focuses on discussing the different issues and challenges impacting the implementation of DFR in SDN environments.

Research by Park, et al. (2018) designed a digital forensic readiness model for a cloud computing-based smart work environment considering the current changes in cloud computing. Their research work was purely based on a cloud computing-based smart work environment and not on issues and challenges impacting the implementation of DFR in SDN environments.

Spiekermann and Eggendorfer, (2016) analysed different challenges in investigating virtual networks. As opposed to the current study in this paper, their research proposed a classification in several different categories to help in developing new methods and possible solutions to simplify investigations in virtual network environments and not necessarily SDN. Karie and Karume, (2017) also presented different issues and challenges surrounding the implementation of digital forensic readiness in organisations. Their research, however, was generic and did not point out any specific implementation of DFR in SDN environments. Though the focus of the current paper is on SDN, the authors acknowledge that some of the different sentiments discussed by Karie and Karume, (2017) are also applicable to SDN environments.

Other related works exist on issues and challenges surrounding DFR in SDN environments, however, neither those nor the cited references in this paper have presented the specific issues and challenges impacting the implementation of DFR in SDN environments in the way that is discussed in this paper. However, the authors acknowledge the fact that the previous research works have offered valuable insights toward the study in this paper. The next section presents a detailed discussion of the different issues and challenges impacting the implementation of DFR in SDN environments.

## 4. Issues and challenges Impacting DFR Implementation in SDN

In this section of the paper, the authors discuss some of the identified issues and challenges impacting DFR implementation in SDN environments. Note that as mentioned earlier this study employed purposeful sampling and for this reason, the issues and challenges identified in this section were only selected to facilitate this study based on the literature sampled by the authors and do not by any means constitute an exhaustive list. Besides, some of the issues and challenges discussed are native to SDN environments hence, more specific issues and challenges can and should be added as technology evolves.

### 4.1 Lack of Standards Focusing on DFR Implementation in SDN.

Like any other existing standard, the primary reason for having internationally recognised DFR standards is "to promote good practise methods and processes for forensic capture and investigation of digital artefacts" (ISO/IEC 27037, 2012). This also implies that standards are generally accepted as good, however, due to the relative newness of SDN infrastructure, as well as having recently gained significant momentum in many organisations (Kandoi and Antikainen, 2015) international standards that focus purely on DFR implementation in SDN environments are yet to be realised. The lack of international standards specifically focusing on DFR implementation in SDN is therefore a challenge that according to Khan et al. (2016), calls for a comprehensive forensic mechanism or standard to help in investigating the different forms of attacks in SDN environments as well as facilitate future forensic investigations.

### 4.2 Lack of DFR Implementation Policies for SDN

To achieve business objectives many organisations, make use of policies. In the context of a legacy network system, policies are a collection of rules, conditions, constraints, and settings defined by network administrators to govern the behaviours of network devices or designate who is authorised to connect to the network and the circumstances under which they can or cannot connect (Microsoft, 2020). In the SDN environment, a DFR implementation policy can thus be understood as a document that details the immediate procedures to be employed to support DFR in an organisation and any future forensic investigation of digital artefacts. Such a policy provides a systematic, standardised, and legal basis for the admissibility of digital evidence that may be required from a formal dispute or legal process (Karie and Karume, 2017). However, correctly setting up and enforcing DFR implementation policies in SDN may require accurate, fine-grained, and trusted information of user applications generating network traffic which may not be available upfront in new SDN environments. A Lack of DFR implementation policies in SDN environments, therefore, makes the development and implementation of any SDN DFR frameworks a very challenging process.

### 4.3 Budget Constraints for Implementing DFR in SDN

Knowing that SDN is still considered relatively new, and that security was not initially a key characteristic of the SDN architecture according to Khan et al. (2016), implementing DFR may be subject to budgetary or cost constraints which may not be known upfront. Moreover, as stated by Reddy, Venter, and Olivier, (2012), a DFR program consists of several activities that should be chosen and managed for the cost constraints and risk of the organisation. However, as organisations opt for cheaper options to be implemented in the place of DFR because of budget constraints, some of the alternatives implemented make digital forensic investigation costly and challenging.

### 4.4 Fear of Network Downtime

When operating on legacy network systems, which in most cases is used to support client business, migrating to SDN means a part of the entire legacy network will have to be shut down for some time causing client business needs to be shut down for the length of the specified downtime as well. This scenario may affect the implementation of DFR in SDN environments especially in an organisation where the process involved cannot be fully automated to facilitate the integration of the legacy network systems into SDN environments while maintaining backward compatibility (QuoteColo, 2016). The fear of network downtime, therefore, is a challenge when an organisation is considering a move to SDN.

### 4.5 Lack of Skilled Personnel

In many business environments, it is expected that the skills offered by the personnel match the skills wanted by organisations to meet their objectives. However, in most cases, this is always not the case. According to Desai et al. (2009), for some time now knowledgeable and skilled digital forensic personnel are hard to find. The shortage of skilled personnel poses a challenge in many different environments including the implementation of DFR in SDN. This, therefore, calls for organisations to address skill shortages to cover existing gaps through training or workshops which can also have other budget or cost implications.

### 4.6 SDN Scalability Challenge

As businesses grow, so is their networking environment. Large SDN environments with volumes of networking requests can overwhelm SDN controllers making the management of information flow between the separate data plane and control plane a challenge to SDN. One possible solution is for organisations to embrace decentralized control architecture as well as more intelligence implemented to the data control plane to intersperse data between multiple control planes. This will also help in monitoring and ensure network device accountability and network latency between connected planes. However, decentralized control architecture can make implementing DFR a challenge to some organisations.

### 4.7 DFR Implementation as a Challenge

Existing standards like the ISO/IEC 27043 are more generic and not application specific according to Kebande, Mudau, Ikuesan, Venter, and Choo, (2020). Compared to legacy network systems, SDN has some level of complexity hence implement SDN technology without reinventing the whole architecture with its aspects and related components can be challenging according to Galis et al., (2013). The lack of standards as well as the complexity of the SDN environment makes implementing DFR a challenge to some organisations.

Considering the newness of SDN and its complexity, other issues and challenges that may also directly or indirectly impact the DFR implementation in any SDN environment are:
- Interoperability,
- Reliability,
- Controller Placement (Controller Bottleneck) and
- Performance.

Note that these issues and challenges are also native to SDN but may in one way or the other have some impact on how DFR is implemented. The next section highlights some of the suggested countermeasures that organisations can embrace to enhance their ability to respond to cybersecurity incidents as well as help them in DFR implementation in SDN environments.

### 5. Suggested Countermeasures to the Issues and challenges Impacting DFR Implementation in SDN

With reference to the issues and challenges identified and discussed in this study, this section offers insights into some of the potential countermeasures that organisations can embrace to enhance their ability to respond to the issues and challenges impacting DFR implementation in SDN environments. However, the authors also acknowledge at this point that the countermeasures discussed in this section are because of the purposeful sampled literature and not in any way a comprehensive list. Research still needs to be done to enhance or add to this list.

## 5.1 Develop DFR Implementation Policies.

Policies are mostly used to set the directional tone for different areas of a business organisation. However, because SDN is still considered new technology, policy development for DFR implementation needs careful planning as well as needs identification. Organisations need to gather as much information before any policy development process. The gathered information helps in drafting and reviewing the policy before the final implementation and subsequent reviews. A Well-defined SDN implementation policy can be a key asset to providing the basis for an organisation to analyse how to get from their existing legacy network systems to having a fully functional forensic-ready SDN environment.

## 5.2 Develop SDN standards.

In this context, the authors believe that, developing internationally accepted SDN standards can help organisations and other industry in applying world best practices. This, therefore, calls for collaboration between business organisations and different industry stakeholders to develop internationally accepted SDN standards that can ease the process of implementing DFR in SDN environments. Standards will also help organisations manage cybersecurity incidents with easy as well as maximize the potential use of digital evidence while minimizing digital forensic investigation costs in SDN environments.

## 5.3 Managing Backward Compatibility and Avoiding Lengthy Downtime.

With the advancement in technology, backward compatibility allows for interoperability with older but existing legacy network systems. This is important because organisations can leverage the power of their new systems while still able to transact using old legacy network systems. For this reason, business organisations should consider, backward compatibility when considering a transition to help reduce the cost associated with network downtown, limit service disruption, and reduce possible network security risks (QuoteColo, 2016) especially during the process of implementing DFR. Besides backward compatibility, creating redundancy as well as preparing an organisation for disaster recovery ahead of time can help avoid lengthy downtimes as redundancy can help increase reliability as well as system performance.

## 5.4 Consider Cost-benefit Analysis for DFR Implementation

With cost-benefit analysis, organisations can estimate the strengths and weaknesses of alternative approaches to implementing DFR in SDN environments. This way organisations can determine what different options exist that can provide the best approach to achieving DFR implementation while managing their budget constraints as well as preserving savings.

## 5.5 Training of Personnel

Both old and new personnel should continually be trained as new technology emerge. Training helps organisation personnel comply with new forensic readiness best practices that help them address existing issues and challenges impacting the implementation of DFR in SDN. Training also ensures that personnel are aware of any existing or new standards, policies, and procedures to be used before, during, and after a digital investigation process. More countermeasures exist beyond what is discussed in this paper and every organisation should consider and explore other existing options including those not mentioned in this study. The next section concludes this paper.

# 6. Conclusion and Future Work

In this paper, the authors have discussed different issues and challenges impacting the implementation of DFR in SDN environments. However, the paper has also highlighted the different countermeasures that organisations can embrace to enhance their ability to respond to cybersecurity incidents as well as help them during the transition process to SDN and more especially during DFR implementation in SDN environments. The presentation in this paper can, for example, help digital forensic practitioners, law enforcement agencies, as well as organisations in developing dynamic and proactive countermeasures to deal with the identified issues and challenges impacting DFR implementation in SDN. However, more research is still needed to provide directions on all the identified issues and challenges as well as the suggested countermeasures to the issues and challenges impacting DFR implementation in SDN. As part of future research, the authors plan to develop a DFR framework that can help ease the implementation of DFR in SDN as well as show how well organisations can deal with some of the identified issues and challenges in this study.

## Acknowledgements

The work has been supported by the Cyber Security Research Centre Limited whose activities are partially funded by the Australian Government's Cooperative Research Centres Programme